\documentclass[usenatbib]{mn2e}
\usepackage{amssymb}

\usepackage{graphicx}
 
\title[SY Cancri, a case for unstable mass transfer?]
{SY Cnc, a case for unstable mass transfer?}

\author[J. Casares et al.]
{J. Casares$^1$, I. G. Mart\'\i{}nez-Pais$^{1,2}$, P. Rodr\'\i{}guez-Gil$^{1,3}$\\
$^1$ Instituto de Astrof\'{\i}sica de Canarias, E-38200 La Laguna, Tenerife, 
Spain\\
$^2$ Departamento de Astrof\'{\i}sica, Universidad de La Laguna, E-38206 La
Laguna, Tenerife, Spain\\ 
$^3$ Isaac Newton Group of Telescopes, Apdo. de Correos 321, E-38700 Santa Cruz
de La Palma, Spain}

\begin{document}

\maketitle

\begin{abstract} 
\noindent  
Intermediate resolution (0.5-1 \AA) optical spectroscopy of the 
cataclysmic variable SY Cnc reveals the spectrum of the donor star. Our
data enable us to resolve the orbital motion of the donor and provide  
a new orbital solution, binary mass ratio and 
spectral classification. 
We find  that the donor star has spectral type G8$\pm$2 V and orbits the white
dwarf with $P=0.3823753\pm0.0000003$ day, $K_2=88.0\pm2.9$ km s$^{-1}$ and 
$V \sin i=75.5 \pm 6.5$ km s$^{-1}$. Our values are significantly different 
from previous works and lead to $q=M_{2}/M_{1}=1.18 \pm 0.14$. This is one
of the highest mass ratios known in a CV and is very robust because it is 
based on resolving the rotational broadening over a large number of metallic 
absorption lines.
The donor could be a slightly evolved main-sequence or descendant from a
massive star which underwent an episode of thermal-timescale mass transfer.
\end{abstract}

\begin{keywords}
stars: accretion, accretion discs -- binaries:close -- stars: individual
(SY Cnc) -- cataclysmic variables
\end{keywords}

\section{Introduction}

Mass transfer in cataclysmic variables (CVs hereafter) is driven by angular 
momentum loss mechanisms. Magnetic braking is considered the main driver for
orbital periods longer than $\simeq$3 h, although its actual prescription 
is under debate \citep[e.g.][]{ivanova03}. As a result of mass transfer, the 
radius of the donor star is adjusted (1) on a thermal time-scale to try 
preserving thermal equilibrium and (2) on a dynamical time-scale to return to 
hydrostatic equilibrium. 
The competition between the changing stellar and Roche lobe radii 
sets the mass transfer rate and hence its stability. For large mass ratios the 
binary could be unstable against rapid mass transfer and a common envelope may
ensue \citep{king88}. 

SY Cnc is one of the brightest Z Cam-type dwarf novae, exhibiting regular
outbursts every $\sim$27 days. \cite{shafter83} reported an orbital period 
$P=0.380 \pm 0.001$ d and $q=M_2/M_1=1.13 \pm 0.35$  which was interpreted as
evidence for unstable mass transfer. Unfortunately, $q$ was derived indirectly 
using observed properties of the $H\alpha$ emission line and this method is not 
entirely reliable because the accretion disc's emissivity is often found to be
non-axisymmetric and contaminated by hot-spots. The best way to constrain $q$ 
is by combining the radial velocity semiamplitude of the donor star $K_2$ with 
the rotational broadening of its absorption lines $V \sin i$ 
\citep{wadehorne88}. 
The issue of the high $q$ value was revisited by \cite{smith05} who measured 
$K_2=127\pm23$ km s$^{-1}$ from a skew map of the Na~{\sc i} IR doublet. However, the 
poor spectral resolution hindered a direct determination of $V \sin i$ and hence 
$K_2$ was combined with $K_1$ 
\citep[from the wings of the $H\alpha$ line,][]{shafter83} to yield a 
revised $q=0.68\pm0.14$. They also classify the donor star as M0, which must 
be substantially evolved in order to fill its 9.1 h Roche lobe. 

Here we present new higher resolution spectra of SY Cnc demonstrating that the 
spectrum of the donor star is clearly detected at optical wavelengths. 
We update the orbital parameters and present the first determination of the 
rotational broadening of the donor star, which removes the lingering 
uncertainty on the true mass ratio. We also find that the spectrum of the donor 
is best fitted by a late G star and discuss the observed properties in 
the light of mass-transfer stability and the evolution of SY Cnc.

\section{Observation and Data Reduction}

SY Cnc was observed over several nights between 1992 and 2008 
using the Intermediate Dispersion Spectrograph (IDS) attached to the 2.5~m 
Isaac Newton Telescope (INT) at the Observatorio del Roque de Los Muchachos. 
The 27 spectra were always obtained in the $H\alpha$ region but using three  
gratings and slightly different instrumental settings (slit widths 
and central wavelengths), which resulted in spectral resolutions in the range 
25-70 km s$^{-1}$. The nights were allocated to various scientific programs 
and hence only a few spectra of SY Cnc were obtained per night. A log of the 
observations is presented in Table \ref{tabobs}. 
Spectra of the stellar templates 61 Cyg A \& B (K5V and K7V) were also 
observed during the 2008 run for the purpose of radial velocities and
rotational broadening analysis. In addition, eight templates of spectral 
types G6-M0V were collected during the 2003 campaign and another K5V in 1999.  

The images were bias corrected and flat-fielded, and the spectra 
subsequently extracted using conventional optimal extraction techniques in
order to optimize the signal-to-noise ratio of the output \citep{Horne86}.  
Every target was bracketed with observations of a comparison CuAr+CuNe arc 
lamp and the pixel-to-wavelength scale
was derived through polynomial fits to a large number of identified
reference lines. The final rms scatter of the fit was always $<$1/30 of
the spectral dispersion.

\begin{table}
\centering
\caption[]{Log of the observations.}
\label{tabobs}
\scriptsize
\begin{tabular}{lcccc}
\hline
\hline
Date    & Wav. Range & Exp. time & Dispersion & Resolution \\
 & $\lambda\lambda$ & (s) &  (\AA\ pix$^{-1}$) &(km s$^{-1}$)\\
\hline
10/03/1992    & 6025-6975  & 2400         & 0.79 & 70 \\
13/02/1998    & 6290-6800  & 600,900  & 0.53 & 49 \\
1-3/05/1999   & 6290-6800  & 2x600,6x900  & 0.53 & 49 \\
22-27/04/2003 & 6100-6780  & 10x600       & 0.30 & 25 \\
20-22/10/2008 & 5575-7139  & 3x300,3x600  & 0.64 & 53 \\
\hline
\end{tabular}
\end{table}

\section{The Orbital Solution}

\begin{figure}
\centering
\includegraphics[width=68mm,angle=-90]{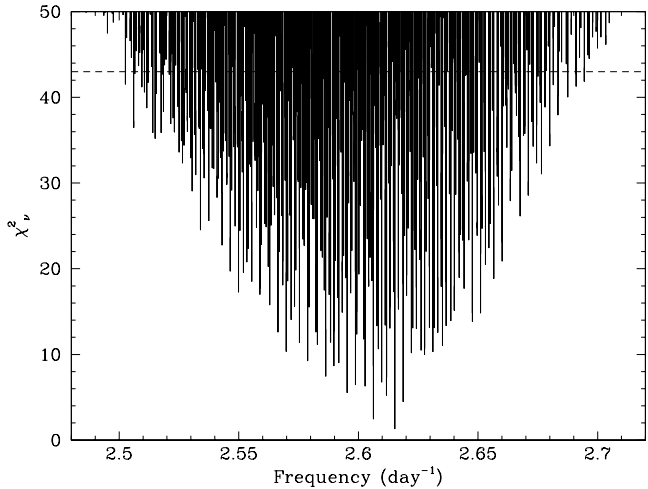}
\caption{The $\chi^2$ periodogram of the radial velocities of the donor star 
around the minimum value. The horizontal dashed line indicates the 
99.99 per cent significance level above noise. The deepest peak corresponds to 
$P=0.3823753$ d.} 
\label{figperiod}
\end{figure}

We rectified the 27 individual spectra by subtracting a low-order spline fit 
to the continuum, after masking out the main emission and atmospheric absorption 
lines. The spectra were subsequently rebinned into a uniform velocity scale of 
36 km s$^{-1}$ pix$^{-1}$. The K5V template 61 Cyg A was broadened to 78 km 
s$^{-1}$ to match the width of the donor photospheric lines (see Sect. 4). 
Every spectrum of SY Cnc was then cross-correlated 
against the broadened template in the spectral regions free from emission and 
telluric absorption features. Radial velocities were extracted following the 
method of \cite{ton79}, where parabolic fits were performed to the peak of the 
cross-correlation functions, and the uncertainties are purely statistical. 
Since the orbital period is poorly constrained to 0.380 $\pm0.001$ d 
we performed a power spectrum analysis on the radial 
velocities in the range 0.1-2 days, and the results are displayed in Fig. 
\ref{figperiod}. Here we have rescaled the errorbars by a factor 1.3 so that 
the minimum $\chi^{2}_{\nu}$ is 1.0. 
The periodogram is dominated by strong aliasing due to the sparse sampling
of our observations. In order to test the significance of the different peaks
above the noise level we performed a Monte Carlo simulation. Synthetic 
$\chi^{2}$ periodograms were computed
from a large population (10$^5$) of velocities randomly picked from a 
white-noise distribution and with identical time sampling as our data. 
A 4-$\sigma$ significance level is defined by the 99.99 per cent of the 
computed $\chi^{2}_\nu$ values and this is indicated in Fig. \ref{figperiod} 
by a horizontal dashed line. Most of the peaks between frequencies 2.5-2.7
day$^{-1}$ are under the line and hence periods in the range 0.37-0.40 days are 
significantly above noise at the 99.99 per cent level.  
The two deeper peaks correspond to 0.3824 day and 0.3837 day and have 
$\chi^{2}_{\nu}=1.0$  and 2.3 respectively for 24 degrees of freedom. 
A 4-$\sigma$ significance level around the minimm peak will exclude the second 
peak \citep{lampton76} and hence we can conclude that 0.3824 day is by far the 
most significant period.
A least-squares sine-wave fit to the radial velocities, using $P=0.3824$
day as input parameter, yields the following parameters

$$ \gamma= -0.2 \pm 2.5~ {\rm km~s}^{-1}$$ 
$$ P=0.3823753 \pm 0.0000003~ {\rm d}$$ 
$$ T_{0} = 2451300.254 \pm 0.002$$ 
$$ K_{2} = 88.0 \pm 2.9~ {\rm km~s}^{-1}$$ 

\noindent 
where $T_{0}$ corresponds to the Heliocentric Julian date of the inferior 
conjunction of the donor star. 
All quoted errors are 68 per cent confidence. The systemic velocity $\gamma$ 
has been corrected from the radial velocity of 61 Cyg A, that we take as 
-64.3 $\pm$0.9 km s$^{-1}$ \citep{wilson53}. 
Note that a sinewave fit fixing $P=0.3837$ day also yields 
$K_2=87.3\pm3.3$ km s$^{-1}$, an indication that our 
$K_2$ value is robust and its error realistic, irrespectively of the true value
of the orbital period. The same is found when other peaks around the minimum are
taken.

\begin{figure}
\centering
\includegraphics[width=68mm,angle=-90]{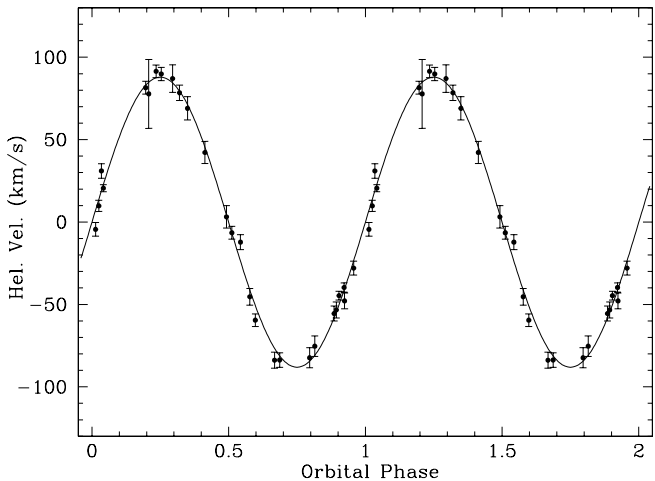}
\caption{Radial velocity curve folded on $P=0.3823753$ day. The
best sine-wave solution is overplotted.}
\label{figrv}
\end{figure}

Fig. \ref{figrv} displays the 
radial velocity points folded on our favoured orbital period $P=0.3823753$ day 
together with the best 
sine fit solution. Our $K_2$ velocity disagrees with the one reported by 
\cite{smith05} which was obtained using the 8190-\AA~ Na~{\sc i} doublet. We 
note that the Na~{\sc i} doublet can be quenched by heating effects, as opposed 
to the metallic lines used by us \citep{martin89}. 
If this were the case, the light centre of the Na~{\sc i} lines would be
displaced towards the back side of the star. The effects of irradiation can be
estimated using the K-correction approach of \cite{wadehorne88} $\Delta
K_2/K_2= f{\bf (}r_2/a{\bf )} (1+q)$, where $\Delta K_2$ is the increment in 
$K_2$ velocity
due to irradiation, $r_2$ the radius of the donor star, $a$ the binary
separation and $f$ the fractional displacement of the absorption line's site
with respect to the center of mass of the star. In the extreme case, when the 
Na~{\sc i} absorption is completely supressed from the irradiated hemisphere, 
$f=4/(3\pi)$. Thus, replacing $r_2/a$ by Eggleton's equation 
\citep{eggleton83} and adopting $q=1.18$ (see next section) we find $\Delta 
K_2=32$ km s$^{-1}$. This difference is just enough to accommodate Smith et 
al's value with
ours, for maximum quenching of the Na~{\sc i} lines. Therefore, we conclude
that Smith et al's $K_2$ is likely a significant overestimate due to
irradiation while our value is much less affected by these effects.

\section{The Rotational Broadening and Spectral Classification}

In order to measure the rotational broadening of the donor's absorption 
features we have focused on the 10 highest resolution spectra gathered 
during the 2003 campaign. We broadened the G6-M0 templates from 5 to 100 
km s$^{-1}$ in steps of 5 km s$^{-1}$, using a Gray profile \citep{Gray92} 
and continuum limb-darkening coefficients appropriate for every spectral 
type and our wavelength range. The broadened templates were multiplied by factors 
$f<1$, to account for the fractional contribution to the total light. These were 
subsequently subtracted from the Doppler corrected average of SY Cnc, obtained 
using our orbital solution above. The optimal broadening, based on a $\chi^{2}$ 
test on the residuals, is always found in the range 77-80 km s$^{-1}$. The mean 
of these determinations is $78.4 \pm 3.5$ km s$^{-1}$. The quoted error
corresponds to 1-$\sigma$ and has been estimated by fitting a cubic
function to the $\chi^{2}$ vs $V \sin i$ curve and searching for 
$\chi^{2}_{\rm min}+1$. 

A potential source of systematics in this calculation 
is the assumption of continuum limb-darkening coefficient because absorption 
lines in late-type stars are expected to have smaller core limb-darkening 
coefficients than the continuum \citep{collins95}. We have tested this by 
repeating the above analysis using zero limb-darkening as a secure lower 
limit and obtain $V \sin i=72\pm3$ km s$^{-1}$. Therefore, we decided to adopt 
$75.5 \pm 6.5$ km s$^{-1}$ as a conservative value for $V \sin i$ hereafter. 
The minimum $\chi^{2}$ also constrains the donor's 
spectral type to G9-K3V. The same analysis was independently performed using the 
lower resolution data from 1999 and 2008, which yield consistent results at 
$80 \pm 5$ and $77 \pm 5$ km s$^{-1}$, respectively. In the latter case the K5V
template produces a lower minimum $\chi^{2}$ than the K7V, an indication that
the donor star is earlier than K5. 

\begin{figure}
\centering
\includegraphics[width=75mm,angle=0]{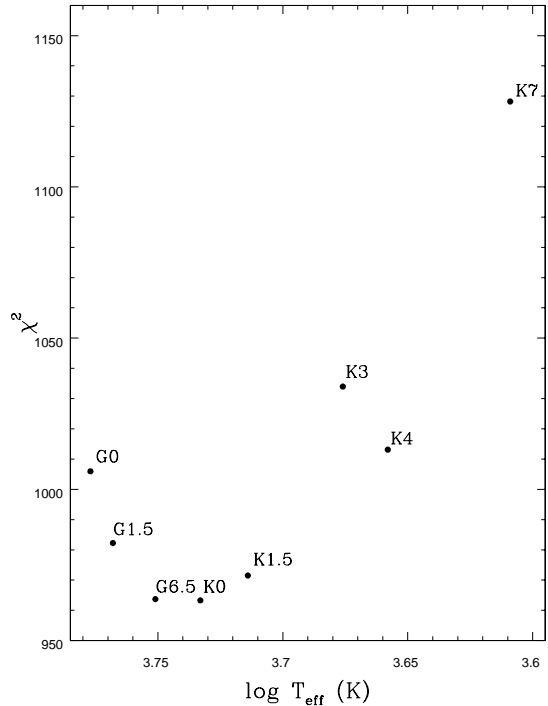}
\caption[]{Spectral type classification of the donor star. The minimum 
$\chi^{2}$ is well constrained between spectral types G6.5V and K0V.}
\label{figsptype}
\end{figure}

\begin{figure}
\centering
\includegraphics[width=68mm,angle=-90]{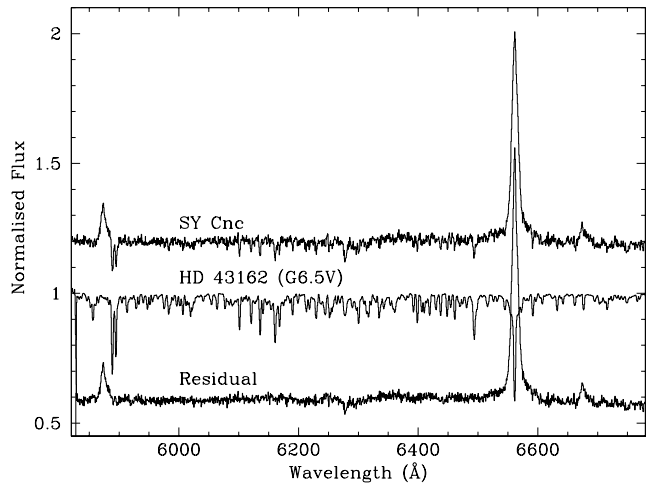}
\caption[]{Doppler corrected average of SY Cnc in the rest frame of the
donor star together with the G6.5V template and the residual after an optimal 
subtraction. The template has been broadened to match the 
$V \sin i$ of the donor star.}
\label{figopts}
\end{figure}

To further constrain the spectral type we employed a set of high 
resolution (R=$\lambda/\Delta\lambda$=55000-11000) spectra covering G0-K7V 
from Ecuvillon et al. (2004, 2006) and \cite{montes98}. These are
HD 39091 (G0V), HD 30495 (G1.5V), HD 43162 (G6.5V), HD69830 (K0V), 
HD 17925 (K1.5V), HD 222237 (K3V), HD 216803 (K4V) and HD 157881 (K7V). 
These stars have 
very accurate stellar parameters $\log g, T_{\rm eff}$ and metallicities 
so they can provide a more robust spectral type classification. 
These were downgraded to the instrumental resolution of the 2008 INT data and 
broadened using our $V \sin i$ estimate. 
Fig. \ref{figsptype} displays the $\chi^{2}$ of 
the optimal subtraction as a function of the template's spectral type. The 
minimum is found for G6.5V and K0V and 
thus we classify the donor star as a G8 $\pm2$V. This is again remarkably 
different from the M0 suggested by \cite{smith05} whereas it is in better 
agreement with the work of \cite{harrison04} who propose an early G star based
on IR spectra. Fig. \ref{figopts} presents the 2008 average spectrum of SY Cnc, 
Doppler corrected in the rest frame of the donor star, together with one of 
the best templates. The latter must be scaled by a factor 0.42$\pm$0.02 to
match the depth of the absorption features in SY Cnc and hence we conclude that
the donor star contributes 42 per cent to the total flux in the $H{\alpha}$ 
region. 

\section{Discussion}

Since the donor star is filling its Roche lobe and synchronized, we can 
combine our determination of $K_2$ and $V \sin i$ to constrain the binary mass
ratio through the equation: 

\begin{equation}
V \sin i = K_{\rm 2}~(1 + q)~\frac{0.49~q^{2/3}}{0.6~q^{2/3}+\ln{(1+q^{1/3})}}.
\end{equation}

\noindent
This uses Eggleton's expression for the effective radius of the Roche lobe 
\citep{eggleton83}, which is more accurate than Pacy\'nski's approximation 
\citep{paczynski71}. Substituting our values of $K_2$ and $V \sin i$ into equation
(1) we find $q=1.18\pm0.14$, where the error has been estimated using a Monte
Carlo simulation. The input parameters $V \sin i$ and $K_2$ have been picked 
up randomly from normal distributions whose mean and sigma correspond to the 
observed values and the process is repeated $10^5$ times. This is one of 
the highest, accurately measured, mass ratio in a CV, only comparable to V363 Aur 
with $q=1.17\pm 0.07$ \citep{thoroughgood04}. For instance, the 7.10 edition of 
the Ritter \& Kolb catalogue \citep{ritter03} lists seven systems in 731 
entries with $q>1$. However, only V363 Aur and AC Cnc possess 
robust $q$ values because these are derived through a measurement of $K_2$ 
and $V \sin i$. The remaining five should be regarded as tentative given the 
difficulty in deriving stellar masses in the absence of donor stellar features. 
 
Theoretical studies of binary evolution show that stable mass transfer
depends on both the thermal and dynamical response of the donor star to mass
loss. This is described by the thermal and adiabatic radius-mass exponents,
computed for the case of ZAMS donor stars \citep[e.g.][]{politano96}. 
When compared with the tidal radius-mass exponent (i.e. the logarithmic 
derivative of the Roche lobe radius with respect to the donor mass) they define 
a critical mass ratio for stable mass transfer.  
The radius-mass exponents, in turn, depend on the donor's mass so a combination 
of ($q,M_2$) values is required to fulfil the condition for stable mass 
transfer.

\begin{figure}
\centering
\includegraphics[width=68mm,angle=-90]{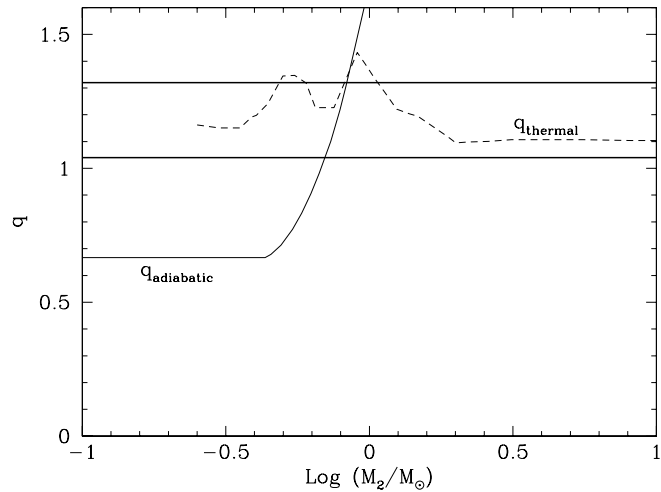}
\caption[]{Binary mass ratio versus donor mass. The solid and dashed lines
represent the critical mass ratio for adiabatic (dynamical) and thermal 
mass-transfer stability, respectively, for ZAMS donor stars. 
The thick horizontal lines mark our limits to $q$.}
\label{figstab}
\end{figure}

SY Cnc presents a remarkable case because of its high (and accurate) mass
ratio which places it close to the absolute upper limit of mass transfer
stability computed by \cite{politano96}. Fig. \ref{figstab} shows the 
critical mass ratio curves for dynamical (solid line) and thermal stability 
(dashed line) as a function of the donor mass, updated after figs. 1 
and 2 from \cite{smith06}. For SY Cnc to lie in the stable 
mass-transfer region (i.e. below the two curves) our central value 
$q=1.18$ yields $M_2\sim $0.77-1.54 M$_{\odot}$. This would imply a 
white dwarf mass $M_1\sim0.65-1.31$ M$_{\odot}$ and, since 
$M_{1} \sin^{3} i  = (1+q)^{2} P K_{2}^{3}/2 \pi {\rm G} = 0.13 \pm 0.02$ 
M$_{\odot}$, then $i\sim26-38^{\circ}$. However, 
the theoretical stability limits are computed under the assumption of ZAMS donor
stars which is not necessarily true. Donor stars in CVs 
have on average later spectral types than field main-sequence stars , an 
indication of radius  
expansion \citep[e.g.][]{knigge06}. In particular, \cite{beuermann98} showed 
that donor stars with $P_{\rm orb}\ga 5-6$ h were likely nuclear-evolved at the 
start of mass transfer. This alters the helium-to-hydrogen composition and 
hence the response of the star to mass loss. Therefore, the critical mass 
ratios shown in Fig. \ref{figstab} should be regarded as mere indicative limits.

Our determination of the spectral type also provides a new ingredient to
constrain the evolution of the donor star in SY Cnc. A G8 ZAMS has $M_2=0.81$ 
M$_{\odot}$ which would place it reasonably well within the stability 
region. A donor with such mass would fill $\simeq87$ per cent of its 
9.1 h Roche lobe. The conclusion seems to be that the companion star is 
a slightly evolved main sequence, in contrast with previous claims by 
\cite{smith05}. 

Alternatively, SY Cnc could be descending from a progenitor binary with 
a more extreme mass ratio which underwent a phase 
of thermal-timescale mass transfer. The importance of this evolutionary path 
has been addressed by several papers 
\citep[e.g.][]{schenker02a, podsiadlowski03, kolb05} and successfully explains 
the cooler-than-main-sequence donor stars typically observed at $\ge5-6$ h 
orbital periods. 
Natural consequences of this scenario are larger white dwarf 
masses and evidence for CNO processed material, both seen in 
AE Aqr \citep{schenker02b}. The detection of nuclear processed material in UV
spectra of SY Cnc would clarify the evolutionary history of SY Cnc and whether it
is the outcome of a thermal mass-transfer episode. Further, more higher 
resolution observations will provide better constraints on $V \sin i$, 
$K_2$, the stellar masses and the evolution of the donor star in this 
important CV.

\section{Acknowledgments}

We thank the anonymous referee for the very helpful comments which improved the 
quality of the manuscript.
MOLLY software developed by T. R. Marsh is gratefully acknowledged. Partly
funded by the Spanish MEC under the Consolider-Ingenio 2010 Program grant 
CSD2006-00070: first science with the GTC. Based on observations made with 
the INT operated on the island of La Palma by the Isaac Newton Group in
the Spanish Observatorio del  Roque de Los Muchachos of the Instituto
de Astrof\'\i{}sica de Canarias (IAC).

\end{document}